\documentstyle[12pt,epsf,amstex]{article}

\begin{document}
\setlength{\unitlength}{1mm}
\textwidth 15.0 true cm 

\headheight 0 cm
\headsep 0 cm
\topmargin 0.4 true in
\oddsidemargin 0.25 true in
\input epsf

\newcommand{\beq}{\begin{equation}}
\newcommand{\eeq}{\end{equation}}
\newcommand{\be}{\begin{eqnarray}}
\newcommand{\ee}{\end{eqnarray}}
\renewcommand{\vec}[1]{{\bf #1}}
\newcommand{\vecg}[1]{\mbox{\boldmath $#1$}}

\renewcommand{\theequation}{\thesection.\arabic{equation}}

\newcommand{\grpicture}[1]
{
    \begin{center}
        \epsfxsize=200pt
        \epsfysize=0pt
        \vspace{-5mm}
        \parbox{\epsfxsize}{\epsffile{#1.eps}}
        \vspace{5mm}
    \end{center}
}

\begin{flushright}

SUBATECH--02--01\\
TPI/MINN-01/51

\end{flushright}

\vspace{0.5cm}

\begin{center}

{\Large\bf  
Born--Oppenheimer corrections to the effective zero-mode Hamiltonian 
in SYM theory.}

\bigskip

   {\Large  A.V. Smilga} \\

\vspace{0.8cm}

{\it SUBATECH, Universit\'e de
Nantes,  4 rue Alfred Kastler, BP 20722, Nantes  44307, France. }\\

\end{center}

\bigskip

\begin{abstract}

We calculate the subleading terms in the Born--Oppenheimer expansion for the
effective zero-mode Hamiltonian of ${\cal N} = 1,\ d=4$ 
 supersymmetric Yang--Mills theory with any gauge group. The Hamiltonian
depends on $3r$ abelian gauge potentials $A_k^s$ ($k = 1,2,3,\ 
 s = 1,\ldots, r$, $r$ is the rank of the group),  and their superpartners.
  The Hamiltonian belongs to the class of 
${\cal N} = 2$ supersymmetric QM Hamiltonian constructed earlier by Ivanov
and I. Its bosonic part describes the motion over the 
$3r$--dimensional
manifold with a special metric. The  corrections explode when the 
root
forms $\alpha_j(A^s_k)$ vanish and the Born--Oppenheimer 
approximation
breaks down.
 
\end{abstract}

\section{Introduction}
The vacuum dynamics of supersymmetric gauge theories has been a 
subject of 
intense interest and study since 1982 when Witten introduced
the notion of supersymmetric index and calculated it for pure 
supersymmetric
gauge theories with unitary and symplectic groups \cite{Witten}.

One of the ways to tackle the problem is to put the system in a 
very small
spatial box and truncate all higher Fourier modes.
The problem is then reduced to a pure quantum mechanical problem, 
which will
be the starting point of our discussion here.

We hasten to comment that such a crude truncation is not quite consistent. A more correct procedure is to integrate out the higher Fourier
modes in the Born--Oppenheimer spirit. This will be discussed in more
details in Sect. 3.
For time being, let us consider a QM system with 2 complex supercharges obtained 
by dimensional reduction from
the original  4--dimensional pure Yang--Mills  ${\cal N} = 1$ 
 supersymmetric  theory  based on the gauge group $G$. This system is called sometimes supersymmetric {\it matrix model} and is interesting on its own.

 The Hamiltonian has the form 
  \be
\label{hamSYM}
\  H \ =\ \frac 12 \  E_i^a  \  E_i^a + \frac {g^2}4 f^{abe} f^{cde}
A_i^a A_j^b A_i^c A_j^d - ig f^{abc} \ {\bar\lambda}^{a\alpha} 
(\sigma_i)_\alpha^{\ \beta}\lambda^b_\beta A^c_i \ , \nonumber \\
i = 1,2,3,\ \ \ \   \alpha = 1,2\ \ \ \ \ \ \ \ a = 1, \ldots , 
{\rm dim}(G) \ .
 \ee
$A_i^a$ are the gauge potentials, $  E_i^a = 
-i\partial/\partial A_i^a$ 
are their canonical momenta operators, and 
$\lambda^a_\alpha$ and 
$\ {\bar\lambda}^{a\alpha} = \partial/\partial \lambda^a_\alpha $
are the fermionic gluino variables and their momenta.
\footnote{The indices are raised and lowered with the $\epsilon$--symbol:
$\psi^\alpha = \epsilon^{\alpha\beta} \psi_\beta, \psi_\alpha 
=  \epsilon_{\alpha\beta} \psi^\beta, \epsilon^{12} = -\epsilon_{12} = 1$.
A Hermitian conjugate of an operator $A_\alpha$ is by definition 
$\bar A^\alpha$. Also, $\psi \chi \equiv 
 \epsilon^{\alpha\beta} \psi_\alpha \chi_\beta, \ \bar\psi \bar\chi \equiv 
 \epsilon_{\alpha\beta} \bar\psi^\alpha \bar\chi^\beta $. }
The Hilbert
space includes only the physical
states annihilated by the Gauss law constraints
 \be
   \label{conSYM}
  \  G^a \Psi \ =\ f^{abc} \left( \  E_i^b A_i^c + i 
\  {\bar \lambda}^{b\alpha} 
\lambda_\alpha^c \right) \Psi \ =\ 0\ .
 \ee

The system has two conserved
complex supercharges
 \be
\label{QSYM}
\  Q_\alpha \ =\  \frac 1{\sqrt{2} }  (\sigma_i)_\alpha^{\ \beta} 
\lambda_\beta^a \left[\  E^a_i -
\frac {ig}2 \epsilon_{ijk} f^{abc} A^b_j A^c_k \right]
 \ee
(they formed a Weyl spinor before reduction) and, being restricted
on the Hilbert space (\ref{conSYM}),  enjoys the ${\cal N} =2$
SQM algebra $\{ \ {\bar Q}^\alpha, \  Q_\beta \}_+ = 
\delta^\alpha_\beta \  H $.
The classical potential in Eq.(\ref{hamSYM}) vanishes in the ``vacuum valleys''
with $f^{abc}A_i^b A_j^c = 0$. Due to supersymmetry, degeneracy
along the valleys survives also after quantum corrections are taken into 
account. As a result, the system tends to escape
along the valleys, the wave function of the low--energy states 
is delocalized, and the 
spectrum is continuous (this implies, incidentally, the continuity of the
mass spectrum of supermembranes \cite{Trieste,deWitt}). 

The vacuum valley (or moduli space) 
is parametrized by $r$ 3--dimensional vectors $A^s_i$ lying
in the Cartan subalgebra ${\mathfrak h}$ of the Lie algebra   ${\mathfrak g}$
of the gauge group $G$.
 As the motion along the valley is infinite, the characteristic values of the
moduli $ A^s_i$ are large. We can subdivide the physical bosonic variables
(there are altogether $3{\rm dim}(G) - {\rm dim}(G) = 2{\rm dim}(G)$ such
variables) into $3r$ slow variables $ A^s_i$ and the fast variables aligned
along the root vectors of   ${\mathfrak g}$. It is natural then to integrate
over the fast variables and to write down the effective Hamiltonian depending
only on the slow variables $ A^s_i$ and their superpartners 
$\lambda_\alpha^s$. To leading order, this Hamiltonian has a rather simple
form \cite{Witten,Trieste,Kac2} :
\footnote{In addition, the requirement of Weyl invariance of the 
wave functions should 
be imposed, but we are not going to discuss it here.}
 \be
\label{Hlaplac}
H^{\rm eff} \ =\ \sum_{s=1}^r \frac 12 E_i^s  E_i^s \ .
 \ee
The corresponding supercharges are
 \be
\label{Qlaplac}
Q_\alpha^{\rm eff} \ =\ \sum_{s=1}^r \frac 1{\sqrt{2}} 
(\sigma_i)_\alpha^{\ \beta} \lambda^s_\beta E_i^s  \ .
 \ee
This paper is devoted to calculation of the subleading corrections to Eqs. 
(\ref{Hlaplac}, \ref{Qlaplac}). We use the method developped earlier
in Ref.\cite{jaSQED} to calculate the corrections to the effective Hamiltonian
in supersymmetric QED. The results are also similar. We will show 
that, for
$SU(2)$ theory, the effective supercharge and the Hamiltonian are given by 
the expressions  
\be
\label{QHN2}
\  Q_\alpha \ =\  \sqrt{\frac 12} \left [ (\sigma_k)_\alpha^{\ \beta} 
\psi_\beta f(\vec{c}) \  P_k + 
i \partial_k f(\vec{c}) 
\  {\bar \psi} \sigma_k \psi  \psi_\alpha  
\right] \ , \nonumber \\
  \  {\bar Q}^\alpha \ =\   \sqrt{\frac 12} \left [\  {\bar \psi}^\beta 
(\sigma_k)_\beta^{\ \alpha} 
f(\vec{c}) \  P_k -
i \partial_k f(\vec{c})  \  {\bar\psi} \sigma_k \psi \  
{\bar \psi}^\alpha  
\right] \ , \nonumber \\
\  H \ = \  \frac 12 f(\vec{c})   P_k^2 f(\vec{c})
 - \epsilon_{jkp} \  {\bar\psi} \sigma_j \psi  f(\vec{c})
\partial_p f(\vec{c})  P_k
- \frac 12  f(\vec{c}) \partial^2_k f(\vec{c})
(  {\bar\psi} \psi )^2 
 \ee
with
 \be
 \label{formf}
f({\bf c}) \ =\ 1 + \frac 3{4g|{\bf c}|^3} \ .
 \ee
Here $c_i \equiv A^3_i$ and $\psi_\alpha \equiv \lambda^3_\alpha$. 
The differential operators $\  P_k = -i\partial/\partial c_k$ and
$\  {\bar\psi}^\alpha = \partial/\partial \psi_\alpha$ 
act on everything on the right they find.

The results (\ref{QHN2}) have  exactly the same form as the 
effective supercharges and Hamiltonian for the photon and photino 
zero modes in dimensionally reduced SQED found in Ref. \cite{jaSQED}. The only 
difference is that the function $f(\vec{c})$ is 
modified: for SQED, the coefficient in the second term in 
Eq.(\ref{formf}) is $-1/4$ instead of $3/4$.  
We will show that for theories based on the groups of higher rank $r$,  the effective Hamiltonian is given by a 
generalization of Eq.(\ref{QHN2}) involving a sum over the roots of
 ${\mathfrak g}$.

\section{Supersymmetric QED.}
\setcounter{equation}0

To make the discussion self--contained, we have to remind briefly the basic
steps of the calculation of $H^{\rm eff}$ in supersymmetric QED.

The theory involves the photon $A_\mu$, the photino $\psi_\alpha$, two Weyl
fermions with opposite charges $\xi_\alpha, \eta_\alpha$ and two oppositely
charged scalars $\varphi, \chi$. 
The charged fields are assumed massless.
We assume also that there is no spatial dependence and
we have quantum mechanics rather than field  theory.
 The dynamical variables $A_i$ are slow and the variables
$\varphi, \chi$ are fast. The supercharges and the Hamiltonian of the system
are convenient to represent as 
 \be
\label{grading}
Q_\alpha \ =\ Q_\alpha^{(0)} + Q_\alpha^{(1)},\ \ \ 
H \ =\ H^{(0)} + H^{(1)} + H^{(2)}\ ,
\ee
where
\footnote{The field $\psi$ is defined here with the extra factor 
$-i$ compared to Ref.\cite{jaSQED}.} 
  \be
\label{Q01}
 Q_\alpha^{(0)}  &=& \left[ - \pi_\varphi \delta_\alpha^\beta + ie 
\bar\varphi A_k (\sigma_k)_\alpha^{\ \beta} \right] \xi_\beta +
\left[ - \pi_\chi \delta_\alpha^\beta - ie \bar\chi A_k 
(\sigma_k)_\alpha^{\ \beta} \right] \eta_\beta  \ ,\nonumber \\
Q_\alpha^{(1)}  &=& \sqrt {\frac 12 }
\left[ P_k (\sigma_k)_\alpha^{\ \beta} - 
i e(\bar\varphi  \varphi - \bar\chi \chi)
\delta_\alpha^\beta\right]\psi_\beta
 \ee
and 
 \be
\label{H012}
H^{(0)} &=& \pi_\varphi \pi_{\bar\varphi} + \pi_\chi \pi_{\bar\chi}
+ e^2 (\bar\varphi \varphi + \bar\chi \chi) A_k^2 + eA_k (\bar \eta \sigma_k \eta
- \bar\xi \sigma_k \xi)\ , \nonumber \\
H^{(1)} &=& e\sqrt{2} (\bar \psi \xi \bar\varphi + \bar \xi \psi\varphi)
- e\sqrt{2}  (\bar \psi \eta \bar\chi + \bar \eta \psi\chi)\ , \nonumber \\
H^{(2)} &=& \frac 12 P_i P_i + \frac 12 e^2 (\bar \varphi \varphi - \bar\chi \chi)^2
  \ee
( $P_k = -i\partial/\partial A_k$, $\pi_\phi = -i\partial/\partial \phi$, etc).
The different terms in Eq.(\ref{grading}) are classified according to 
the powers of the small parameter $|\varphi|/|\vec{A}| \sim   |\chi|/|\vec{A}|$.
The leading order Hamiltonian $H^{(0)}$ is quadratic with respect to the fast
variables $\phi, \chi$ and their superpartners $\xi_\alpha, \eta_\alpha$ and
represents a supersymmetric oscillator. The ground state of $H^{(0)}$
has zero energy and its wave function can be easily found:
 \be
\label{vacH0}
\phi_0(\varphi, \bar\varphi, \chi, \bar\chi; \xi_\alpha, \eta_\alpha) \ =
\nonumber \\ 
\frac e\pi A \exp\{-eA(\bar\varphi\varphi + \bar\chi\chi) \}
\left[\xi^\alpha \eta_\alpha + \xi^\alpha (\sigma_k)_\alpha^{\ \beta} 
\eta_\beta A_k/A 
\right]  \ .
 \ee
 Here 
$A_k$ enter as  parameters ( $A = |\vec{A}|$). The characteristic values of  
 $\varphi, \chi$ in the wave
function (\ref{vacH0}) are $\varphi_{\rm char} \sim \chi_{\rm char} \sim 1/\sqrt{eA}$.
We see that the assumption $A \gg |\varphi|, |\chi|$ is self--consistent when 
$\gamma = 1/(eA^3) \ll 1$. Now, $\gamma$ is the actual Born--Oppenheimer parameter
in $H^{\rm eff}$, the corrections over which we are going to find.
  
To do this, we have to represent the total wave function of our system as a sum over 
the eigenfunctions of  $H^{(0)}$:
  \be
\label{expan}
\Psi(A_i,\psi_\alpha; \varphi, \bar\varphi, \chi, \bar\chi, \xi_\alpha, \eta_\alpha) = 
 \sum_n r_n(A_i, \psi_\alpha)  
\phi_n(\varphi, \bar\varphi, \chi, \bar\chi; \xi_\alpha, \eta_\alpha) \ .
 \ee
Then we write the Schr\"odinger equation  $H\Psi = E\Psi$, assuming the energy $E$
being small compared to the characteristic energies $E_n$ of the excitations of
$H^{(0)}$, express the coefficients
$r_{n>0}$ via $r_0$ and cast the equation for $r_0$ thus obtained in the 
Schr\"odinger
form. The operator acting on $r_0$ {\it is} called the effective Hamiltonian.

It is technically convenient to calculate the effective supercharge rather than the 
effective Hamiltonian. To leading order, the former is given by the matrix element
\footnote{When deriving this, we used the fact that $\langle \bar\varphi \varphi \rangle_{00}
= \langle \bar\chi \chi \rangle_{00} $ and also that the derivative of the wave function
(\ref{vacH0}) with respect to $A_i$ has no projection on $\phi_0(\varphi, \bar\varphi, 
\chi, \bar\chi; \xi_\alpha, \eta_\alpha)$. }
 \be
\label{Qeff0}
Q_\alpha^{\rm eff} = \langle Q_\alpha^{(1)} \rangle_{00} 
\ =\ \sqrt{\frac 12} (\sigma_k)_\alpha^{\ \beta} \psi_\beta P_k\ .
 \ee
Taking the subleading correction into account, we obtain [see Eqs.(15,16) 
of Ref.\cite{jaSQED}]
 \be
\label{Qeffpopr}
Q_\alpha^{\rm eff} = \langle Q_\alpha^{(1)} \rangle_{00} -
\sum_n' \frac{\langle Q_\alpha^{(1)} \rangle_{0n}\langle H^{(2)} 
\rangle_{n0}}{E_n}
+ \sum_{nm}'  
 \frac{\langle Q_\alpha^{(1)} \rangle_{0n}\langle H^{(1)} \rangle_{nm}
\langle H^{(1)} \rangle_{m0}}{E_n E_m} \ ,
 \ee
where the sums are done over the excited states of $H^{(0)}$. An explicit 
calculation described in  Appendix A  gives us  the first line in 
Eq.(\ref{QHN2}) with
 \be
\label{fSQED}
f(\vec{A}) = 1 - \frac 1{4eA^3}\ .
  \ee  
The second line is obtained from the first by Hermitian conjugation 
and the effective Hamiltonian is calculated as
the anticommutator $\frac 12 \left\{\bar Q^\alpha,  Q_\alpha \right\}_+$.

The bosonic part of the Hamiltonian (\ref{QHN2}) describes the motion along a 
3--dimensional manifold with the conformally flat metric $ds^2 =
f^{-2}(\vec{A}) d\vec{A}^2$. The whole Hamiltonian represents a nonstandard ${\cal N}=2$ 
supersymmetric
extension of such  $\sigma$-model (the standard one involves three complex 
fermionic variables $\psi_i$ instead of two--component $\psi_\alpha$ and
enjoys only ${\cal N}=1$ supersymmetry). This nonstandard extension can be
constructed only in the  quantum mechanical
limit and, in contrast to the standard one, does not allow for a field theory
generalization.

The model  (\ref{QHN2}) can be expressed in superfield (or better to say, 
supervariable) form \cite{Ivanov}. Let us take a usual vector superfield
$V(t, \theta_\alpha, \bar\theta^\alpha)$ involving photon and
photino variables (spatial dependence is suppressed and
no distinction between undotted and dotted indices is made). Introduce the
real symmetric tensor supervariable 
  \be
 \label{phiAB}
\Phi_{\alpha\beta} \ =\ (D_\alpha \bar D_\beta + D_\beta \bar D_\alpha) V\ ,
 \ee
where $D_\alpha, \bar D_\alpha$ are supersymmetric covariant derivatives.
Remarkably, $\Phi_{\alpha\beta}$ is gauge invariant in the QM limit. The 
 real supervariable \\
 $\Phi_k = -(1/4) \epsilon^{\beta\gamma}
(\sigma_k)_\gamma^{\ \alpha} \Phi_{\alpha\beta}$ can be decomposed into 
components as follows:
 \be
\label{Phii}
\Phi_k = A_k + \bar{\tilde\psi} \sigma_k \theta + 
\bar\theta \sigma_k \tilde\psi + 
\epsilon_{kjp} \dot A_j \bar\theta \sigma_p \theta + D\bar\theta 
\sigma_k \theta
\nonumber \\
+ i (\bar\theta \sigma_k \dot{\tilde\psi} - \dot{\bar{\tilde\psi}} 
\sigma_k \theta)
\bar\theta \theta
+ \frac 14 \ddot A_k
 \theta^2 \bar\theta^2
 \ee
($D$ is the auxilliary field). The real symmetric supervariable
 $\Phi_{\alpha\beta}$ satisfies the constraint
  \be
\label{constPhi}
D_\alpha \Phi_{\beta\gamma} + D_\beta \Phi_{\alpha\gamma} + 
D_\gamma \Phi_{\alpha\beta}  = 0
 \ee
and can be {\it defined} via this constraint. 

Consider the action
 \be
\label{SFPhi}
S\ =\ \int dt d^2\theta d^2 \bar\theta \ F(\vecg{\Phi})\ ,
 \ee
where $F$ is an arbitrary function. Expanding the action (\ref{SFPhi}) into
components, we obtain
 \be
\label{Lcomp}
{\cal L} = \frac 1{2f^2} \dot A_j  \dot A_j + \frac i{2f^2}
\left(\bar{\tilde \psi} \dot{\tilde\psi} - \dot{\bar{\tilde\psi}}\tilde\psi 
\right) - \frac {\partial_m f}{f^3} \epsilon_{mjp}  \dot A_j 
\bar{\tilde\psi} \sigma_p \tilde\psi + \nonumber \\
\frac {D^2}{2f^2} + \frac {D \partial_k f}{f^3} \bar{\tilde\psi} 
\sigma_k \tilde\psi - \frac 18 \partial^2 \left( \frac 1{f^2} \right) 
\bar{\tilde\psi}^2 \tilde\psi^2\ ,
 \ee
where we introduced
 \be
\label{fF}
 f(\vec{A}) = \sqrt{\frac 1{2\partial^2 F(\vec{A})}}\ .
   \ee
It is convenient to make the substitution $\tilde\psi = f(\vec{A}) \psi $.
Then $\bar\psi$ and $\psi$ are canonically conjugated.
After eliminating the auxilliary field $D$ and deriving the Hamiltonian
by the standard rules, one can be easily convinced that it identically 
coincides with the Hamiltonian (\ref{QHN2}).    
In our particular case,
 \be
 \label{vidF}
F\ =\ \frac {\vecg{\Phi}^2}{12} - \frac 1{4e|\vecg{\Phi}|} \ln |\vecg{\Phi}|\ .
 \ee
The model (\ref{SFPhi}) can be easily generalized by introducing a set of
$r$ real supervariables $\vecg{\Phi}^{s}$ satisfying the constraints 
(\ref{constPhi}). The action is written as
 \be
\label{SFPhim}
S\ =\ \int dt d^2\theta d^2 \bar\theta F(\vecg{\Phi}^s)
 \ee
with arbitrary $F$. The bosonic part of (\ref{SFPhim}) describes the motion
on a $3r$--dimensional manifold with the metric
  \be
\label{metricm}
G_{is, jt} \ =\ 2 \delta_{ij} \frac{\partial^2 F(\vec{A}^{1},
\ldots, \vec{A}^{r}) }{\partial A_l^{s} \partial A_l^{t}}
  \ee
($i,j = 1,2,3$ and $ s,t = 1,\ldots,r$).

The main result of this paper is that the effective action for  the SYM
quantum mechanics (\ref{hamSYM}) with any gauge group can be cast in the 
form (\ref{SFPhim}) with some particular function $F$ depending on $r$ (the 
rank of the group)  supervariables $\vecg{\Phi}^s$ described above.

\section{$SU(2)$ theory.}

The simplest non--Abelian model of the class (\ref{hamSYM}) is based on the $SU(2)$
gauge group. 
To find  effective theory, we choose 
the  $A_k^3 \equiv c_k$ and its superpartner 
$\lambda_\alpha^3 \equiv \psi_\alpha$ as the slow variables.
The components $A_k^{1,2}$ will be treated as fast variables
  and  integrated over. Heuristically, the variables 
$A_k^1 \pm iA_k^2$ play the same role as the charged scalar fields
$\phi, \chi$ in SQED. This is especially suggestive if the gauge
  \be
\label{gfSU2}
  c_k A_k^{1,2} = 0
 \ee
 is imposed 
--- the number of remaining degrees of freedom would be
just correct. In addition, if aligning $c_k$ along the third
axis and introducing
 \be
\label{fihi}
\phi &=& \frac 14 \left(A_1^1 - iA^2_1 + i A_2^1 + A_2^2 \right)
\nonumber \\
\chi  &=& \frac 14 \left(A_1^1 + iA^2_1 + i A_2^1 - A_2^2 \right)\ ,
   \ee
the potential part of the Hamiltonian has the same form as in SQED.
The analogy should not, however, be pushed too 
far. Though the components $A_k^{1,2}$ and their superpartners
play qualitatively the same role as the charged scalars and 
fermions in SQED, the total Hamiltonian  in the non--Abelian
case is different.

A direct calculation of the effective supercharges and Hamiltonian
along the same lines as it was done for SQED is not a simple task.
The symmetry arguments dictate us, however, that the {\it functional
form} of the effective Hamiltonian is given, again,  by Eq.(\ref{QHN2}) 
(this is the only known ${\cal N} =2$
supersymmetric structure involving three bosonic and two
 complex fermionic variables). The only question is what the 
function $f(\vec{c})$ is.

One thing can be said immediately: $f(\vec{c})$ 
should tend to 1 when
$\vec{c}$ tends to infinity: in this limit 
the corrections in the Born--Oppenheimer expansion vanish. 
The parameter of the expansion is, again, $1/(g|\vec{c}|^3)$.
If this parameter is large,  the fluctuations 
of the fast variables $[A_i^{1,2}]^{\rm char} \sim 
1/\sqrt{g|\vec{c}|} $ are small compared to $|\vec{c}|$.
And if not, then not. This suggests
 that
  \be
\label{fC}
f(\vec{c}) \ =\ 1 + \frac C{g|\vec{c}|^3} + \ldots\ ,
  \ee
which is similar to what we have found above for SQED, and only
the coefficient $C$ may be different.

To determine $C$, let us go back to the Abelian case and 
consider a little bit more sophisticated problem of finding
the effective Hamiltonian not for the dimensionally reduced SQED
(\ref{Q01}, \ref{H012}), but for the 4-dimensional field theory 
put in a small spatial box of length $L$. 
The system involves now an infinite
number of dynamic variables --- the Fourier harmonics of the fields
$A_i(\vec{x}), \ \phi(\vec{x}),\  \xi(\vec{x})$, etc. 
However, the effective theory is written in terms of
a finite number of slow variables, the
zero Fourier harmonics $\vec{A}^3 \equiv \vec{c}/L$ 
of the gauge fields  and their 
superpartners. The effective quantum--mechanical Hamiltonian
has  the same form as in Eq.(\ref{QHN2}) (up to the dimensional
factor $1/L$) with the function
$f(\vec{c})$ given by the expression \cite{jaSQED} 
  \be
\label{sumharm}
f^{\rm field\ theory}(\vec{c}) = 
1 - \frac {e^2}4 \sum_{\vec{n}} 
\frac 1{|e\vec{c} - 2\pi \vec{n}|^3}\ .
 \ee
The sum runs over all 3-dimensional integer $\vec{n}$.
This sum diverges logarithmically at large $|\vec{n}|$. The origin
of the divergence is very clear --- it is just the charge 
renormalization. Substituting Eq.(\ref{sumharm}) into 
Eq.(\ref{Lcomp}) (multiplied by $L$), keeping track of only 
logarithmically divergent part, and introducing  the running 
effective charge
 \be
\label{runcharg}
e^2(L) \ =\ e_0^2 \left[1 - \frac {e_0^2}{4\pi^2} 
\ln(\Lambda_{UV}L) \right]\ ,
 \ee
we obtain the familiar renormalization of the kinetic term in
the effective lagrangian, 
 \be
\label{Leffrun}
 {\cal L}^{\rm eff} \ =\ \frac L2 
\frac {e_0^2}{e^2(L)} \dot{\vec{c}} \dot{\vec{c}} \ + \ldots 
 \ee
 Multiplying ${\cal L}^{\rm eff} $ by the $Z$ factor, one can obtain the renormalized effective Lagrangian with
  \be
\label{freg}
f_{\rm reg}(\vec{c}) = 
1 - \frac {e^2}4 \left[ \sum_{\vec{n}} 
\frac 1{|e\vec{c} - 2\pi \vec{n}|^3} - 
\sum'_{\vec{n}}  \frac 1 {|2\pi \vec{n}|^3} \right] \ .
 \ee

Note that the coefficient $1/4$ in Eq.(\ref{fSQED}) 
{\it implies} the 
correct value $b_0 = -2$ of the one--loop coefficient of the 
$\beta$--function in SQED. If the same program is carried out
for the $SU(2)$ theory, the coefficient in Eq.(\ref{Leffrun})
should depend on the non-Abelian effective charge with $b_0 = 
3c_V = 6$. Therefore, the second term in Eq.(\ref{sumharm}) acquires
the factor $-3$ in the non--Abelian case.
 \footnote{A similar sum, but 
without taking into account the dependence of the metric on 
$\vec{c}$ was written in recent \cite{vanBaal}. See also
\cite{Lusch} for a related analysis in pure Yang--Mills theory.}
This fixes the coefficient $C = 3/4$ in Eq.(\ref{fC})
and leads us to the result (\ref{formf}). This conclusion is
confirmed by an explicit calculation of $f(\vec{c})$ 
in the Lagrange formalism
\cite{Akhm}.

\section{Other Gauge Groups.}
\setcounter{equation}0

Let us first remind some basic facts of the theory of Lie
algebras (see e.g. the textbook \cite{Zhelob}).
 The generators of an arbitrary Lie algebra ${\mathfrak g}$ form 
a linear space formed by $r$
commuting generators of the Cartan subalgebra ${\mathfrak h}$, 
the positive root vectors $e_{\alpha_j} 
\equiv e_j$ and the negative root vectors  $e_{-\alpha_j} \equiv 
f_j$. The relations
  \be
\label{roots}
[h, e_j] \ =\ \alpha_j(h) e_j,\ \ \ \ [h, f_j] \ =\ -\alpha_j(h) 
f_j\ 
 \ee
hold, where $h \in {\mathfrak h}$ and $\alpha_j(h)$ are certain 
linear forms on the Cartan subalgebra
called the (positive) {\it roots} of the Lie algebra 
${\mathfrak g}$. The commutator $[e_j, f_j]$
is proportional to the {\it coroot} 
$\alpha^\lor_j$ lying in  ${\mathfrak h}$. We can choose
 the normalization with
 $[e_j, f_j] =  \alpha_j^\lor$.

Setting a natural metric on  ${\mathfrak h}$ (with 
$\langle h,g \rangle = 
{\rm Tr} \{hg\}$, \ 
$h,g \in {\mathfrak h}$) and the induced
metric on the space of roots, one can define
  the matrix of the scalar products of the roots
$c_{jj'} = \langle \alpha_j, \alpha_{j'} \rangle$. It 
is related to the
so called  {\it Cartan matrix}.
\footnote{The latter is usually defined as the matrix of scalar 
products of {\it simple} roots, whereas in our case 
$\alpha_j$ represent 
all positive roots. To be as clear and instructive as possible, 
we spell out in  Appendix B all the notations for the
simplest nontrivial case of $SU(3)$.}
We normalize $d_j = c_{jj} = 1$ for the long roots. For    
 $Sp(2r)$ and $SO(2r+1)$ there are also short roots 
with $d_j = 1/2$.
The short root of $G_2$ has length $d_j = 1/3$.

The  coroots corresponding to the short roots are long,
Tr$\{\alpha_j^\lor \alpha_j^\lor \} \propto 1/d_j$. The following
corollary is important for us. Consider
the set of  generators $S_j = (e_j, f_j, \alpha_j^\lor)$ 
realizing an embedding 
of $su(2)$ into ${\mathfrak g}$. The Yang--Mills action
$ -\frac 1{2g^2} {\rm Tr} \{F_{\mu\nu} F_{\mu\nu} \}$
restricted on the set $S_j$ gives us the Yang--Mills action for
 the $SU(2)$ group with the effective coupling constant
 \be
\label{gj}
g^{(j)} \ =\  g\sqrt{d_j}\ .
 \ee   
We will also use the  property 
 \be
\label{sumcal}
 \sum_{j'} c_{jj'} \alpha_{j'}(h) \ =\ \frac {c_V}2 \alpha_j(h)\ ,
 \ee
where $c_V$ is the adjoint Casimir eigenvalue. A related 
property is
\be
\label{sumroot}
\sum_j \alpha_j(X^s) \alpha_j(Y^s) \ =\ \frac {c_V}2 
\sum_s X^s Y^s\ . 
 \ee

The effective Hamiltonian of the theory (\ref{hamSYM}) with an arbitrary
gauge group $G$ is expressed in terms of the slow variables $A_k^s$ lying in  ${\mathfrak h}$
 and their superpartners. 
 \footnote{An implicit assumption here is that classical vacua $F_{ij} = 0$ are given by constant gauge potentials lying in  ${\mathfrak h}$. 
For orthogonal and exceptional groups, also nontrivial flat connections exist and are relevant \cite{Kac1}. This gives extra disconnected components in moduli space, but we will not consider this issue here.}
 The other components of the 
vector potential $A^{+j}_k$ and
 $A^{-j}_k$ directed along the root vectors $e_j$ and $f_j$, respectively [for $SU(2)$, 
$A^\pm_k = (A^1_k \pm iA^2_k)/2$ ], represent fast variables. 
We can impose the following gauge fixing condition, 
 \be
\label{kalfiks}
A_k^{\pm j} \ \alpha_j \left(A_k^s\right)\ =\ 0\ ,
  \ee
where $\alpha_j \left(A_k^s\right)$ are the root linear forms
 of the arguments $ A_k^s$.
[$ \alpha_j \left(A_k^s\right)$ is not just a number like $\alpha_j(h)$ is, but a 3--vector.]
Eq.(\ref{kalfiks}) involves $d-r$ gauge fixing conditions ($d$ is the dimension of the group and
$r$ is its rank). After such a partial gauge fixing, the gauge group is broken down to its maximal
torus $[U(1)]^r$ and the problem is reduced to analyzing an Abelian  $[U(1)]^r$ theory. 
  Due to the
conditions (\ref{kalfiks}),  $A^{+j}_k$ and $A^{-j}_k$ involve $2 + 2$ independent components.
If aligning  $\alpha_j \left(A_k^s\right)$ along the third spatial axis, these four components
can be traded for two complex ones
 \be
\label{fihij}
\varphi^j \ =\ \frac 12 \left( A^{-j}_1 + i A_2^{-j} \right),\ \ \ \ 
\chi^j \ =\ \frac 12 \left( A^{+j}_1 + i A_2^{+j} \right) \ . 
  \ee
The fields $\varphi^j, \chi^j$ are charged with respect to the
 set of $r$ Abelian gauge fields 
$A_k^s$. The same charges are carried by the fermion fields
 \be
\label{XiEtj}
\xi_\alpha^j = \sqrt{2}\lambda^{-j}_\alpha \ ,\ \ \ \ \ \  
\eta_\alpha^j  = \sqrt{2}\lambda^{+j}_\alpha \ .
 \ee
The quadratic in fast
variables  part of the Hamiltonian is 
 \be
\label{Hfast2}
\sum_j \left\{ \pi_{\phi^j} \pi_{\bar\phi^j} + \pi_{\chi^j} 
\pi_{\bar\chi^j} +    
g^2 (\bar\varphi^j \varphi^j + \bar\chi^j \chi^j) 
\left[\alpha_j(A_k^s)\right]^2 +  \right. \nonumber \\
\left. 
g\alpha_j(A_k^s) (\bar \eta^j \sigma_k \eta^j
- \bar\xi^j \sigma_k \xi^j)\right\}\ ,
 \ee
which is rather analogous to Eq.(\ref{H012}).
Unfortunately, the full Hamiltonian of the Abelian theory obtained
after a partial gauge fixing (\ref{kalfiks}) is not simple. 
We are not going to tackle it explicitly, but rather reconstruct
the effective Hamiltonian
 from symmetry considerations, as we did for $SU(2)$.
 
The effective Hamiltonian enjoys ${\cal N} =2$ supersymmetry and 
involves $3r$ bosonic
and $2r$ complex fermionic variables. In addition, it involves at 
most quadratic terms
in momenta. The only known candidate theory has the 
form (\ref{SFPhim}), 
and the only question is what is the function $F(\vecg{\Phi}^s)$.
To the leading Born--Oppenheimer order, it is quadratic in 
$\vecg{\Phi}^s$,
 \be
\label{Ffree}
F(\vecg{\Phi}^s) \ =\   \frac 1{12}  \sum_s
\left(\vecg{\Phi}^s\right)^2 \ =\   \frac 1{6c_V}  \sum_j
\left(\vecg{\Phi}^{(j)}\right)^2\ ,
 \ee 
where we introduced the notation $R^{(j)} = \alpha_j(R^s)$ 
and used the relation (\ref{sumroot}). Integrating (\ref{Ffree})
over $d^4\theta$, we obtain
 \be
\label{Lfree}
{\cal L}^{\rm eff} \ =\ \frac 12 \sum_s \left[ \dot{A}_k^s \dot{A}_k^s 
+ i \left( \bar\psi^s \dot{\psi^s} - 
\dot{\bar\psi^s} \psi^s \right) \right]
 \ee
in accordance with Eq.(\ref{Hlaplac}). 

 We are set to calculate
the leading nontrivial Born--Oppenheimer corrections to 
${\cal L}^{\rm eff}$. Born--Oppenheimer approximation is valid
when the fluctuations of the fast variables $\varphi^j, \chi^j$  
are much less than the characteristic values of $|\vec{A}^s|$.
It is clear from  Eq.(\ref{Hfast2}) that 
 $ (\varphi^j)^{\rm char}, (\chi^j)^{\rm char}$ are of order 
$1/g\sqrt{|\vec{A}^{(j)}|}$, which should be compared with 
$|\vec{A}^{(j)}|$. Thereby the Born--Oppenheimer expansion makes
sense when {\it all} the parameters 
$$\gamma^j \ =\ \frac 1{g |\vec{A}^{(j)}|^3}$$
are small. The leading corrections to ${\cal  L}^{\rm eff}$ and,
in particular, to the metric should be linear in $\gamma^j$. This
means that the leading nontrivial correction
to the leading order result (\ref{Ffree}) for the function 
$F(\vecg{\Phi}^s)$ represents a linear form of 
$\delta^j = (1/|\vecg{\Phi}^{(j)}|) \ln |\vecg{\Phi}^{(j)}|$
[cf. Eq.(\ref{vidF})].
In other words,
  \be
\label{Fsumj}
F(\vecg{\Phi}^s) \ =\   \sum_j \left[  \frac 1{6c_V}
\left(\vecg{\Phi}^{(j)}\right)^2 +  
\frac {C_j}{|\vecg{\Phi}^{(j)}|} \ln |\vecg{\Phi}^{(j)}|\right]
\ ,
 \ee 
where $C_j$ are some numerical coefficients, to be determined.
 Eq.(\ref{metricm})   provides us with the metric 
  \be
\label{metric}
 G_{kp, st}  = \delta_{kp} \frac 1{c_V} \sum_j \alpha_j^s \alpha_j^t \left[1 -
\frac {C_jc_V}{g|\vec{A}^{(j)}|^3}  \right]\ .
  \ee
This gives us the kinetic part in the effective Lagrangian
 \be
\label{Lkin}
{\cal L}^{\rm eff} _{\rm kin} \ =\ \frac 1{c_V} \sum_j \dot{\vec{A}}^{(j)}
  \dot{\vec{A}}^{(j)} \left[1 -
\frac {C_jc_V}{g|\vec{A}^{(j)}|^3} \right]\ .
 \ee

Let us consider the situation when one of the root forms
$|\vec{A}^{(j_0)}|$ is much smaller than all others (and the
corresponding parameter $\gamma^{j_0}$ is much bigger than
all others, but still small). Then one
can neglect all the corrections except the one with $j=j_0$ and 
write 
   \be
\label{Lkin0}
{\cal L}^{\rm eff} _{\rm kin} \ =\ \frac 1{c_V}
\left\{\dot{\vec{A}}^{(j_0)}
  \dot{\vec{A}}^{(j_0)}  \left[1 -
\frac {C_{j_0}c_V}{g|\vec{A}^{(j_0)}|^3} \right] +
 \sum_{j\neq j_0} \dot{\vec{A}}^{(j)} 
\dot{\vec{A}}^{(j)} \right\}\ .
 \ee
The relevant dynamics is determined by the variables
$\vec{A}^{(j_0)}$ and their superpartners. It would not be correct,
however, just to cross out the terms with $j \neq j_0$. The 
variables $\vec{A}^{(j \neq j_0)}$ have nonzero projections on 
$\vec{A}^{(j_0)}$:
$$  \vec{A}^{(j)}\ =\  \frac{c_{jj_0}}{d_{j_0}} 
\vec{A}^{(j_0)} +\ {\rm orthogonal\ 
combinations.}$$
Substituting this in Eq.(\ref{Lkin}) and using the property
$$\sum_j c_{jj_0}^2 = \ d_{j_0}\frac{c_V}2 \ ,$$ 
which is a corollary of 
Eq.(\ref{sumroot}), we obtain
  \be
\label{Lkin01}
{\cal L}^{\rm eff} _{\rm kin} \ \approx\ \frac 1{2d_{j_0}}
\dot{\vec{A}}^{(j_0)}
  \dot{\vec{A}}^{(j_0)}  \left[1 -
\frac {2 d_{j_0}
C_{j_0}}{g|\vec{A}^{(j_0)}|^3} \right] +
 {\rm orthogonal\ terms.}
 \ee
For long roots, $d_{j_0} = 1$ and  we can directly compare 
 this with the effective Lagrangian of the $SU(2)$
theory [see Eq.(\ref{formf})]. We obtain
 \be
\label{Bj}
 C_j \ =\ \frac 3{4} \ .  
 \ee
For short roots, we have first to renormalize 
$ \vec{A}^{(j_0)} \to  \sqrt{d_{j_0}} \vec{A}^{(j_0)}$ to 
bring the kinetic term into the standard form. Identifying
the action (\ref{Lkin01}) with that of the corresponding
$SU(2)$ theory and  
bearing in mind that the $SU(2)$ coupling constant 
is related to $g$ according to (\ref{gj}), we derive that
the relation (\ref{Bj}) holds also for the short roots.

Substituting Eq. (\ref{Bj}) into  the Lagrangian 
(\ref{Lkin}), calculating the canonical momenta, and 
evaluating the supercharges with the N\"other theorem, we
obtain   
  \be
\label{Qeffjtot}
Q_\alpha^{\rm eff} \ =\ \frac {\sqrt{2}}{c_V}  \sum_j  \left\{  
(\sigma_k)_\alpha^{\ \beta}  \psi_\beta^{(j)} 
P_k^{(j)}
\left[ 1 + \frac {3c_V}{8g|\vec{A}^{(j)}|^3 } \right] \right. - 
\nonumber \\ 
\left.  
 \frac{9i c_V A_k^{(j)}}{8g|\vec{A}^{(j)}|^5} \bar\psi^{(j)}
\sigma_k \psi^{(j)} \psi_\alpha^{(j)} \right\} 
+ \ldots 
   \ee
where the dots stand for the terms
of higher order in $\gamma^j$. 
\footnote{These terms are important to close the algebra. 
If only the
terms explicitly displayed in Eq.(\ref{Qeffjtot}) 
were taken into account, the property 
$\{Q_\alpha, Q_\beta\}_+ = 0$ would not hold.}
The effective Hamiltonian is
   \be
\label{Heffjtot}
H^{\rm eff} \ =\ \frac 1{c_V} \sum_j \left\{ \left[ \vec{P}^{(j)} \right]^2
\left[1 + \frac {3c_V}{4g|\vec{A}^{(j)}|^3}\right] -
\frac{9c_V A_p^{(j)} }{8g|\vec{A}^{(j)}|^5 }
\epsilon_{lkp} \bar\psi^{(j)}\sigma_l \psi^{(j)} P_k^{(j)} \right. 
\nonumber \\ -
\left. \frac{9c_V}{8g|\vec{A}^{(j)}|^5} \left( \bar\psi^{(j)} 
\psi^{(j)} \right)^2   \right\} +\ 
 \ldots
 \ee

If the original non--Abelian  theory is placed in a
finite spatial box, one has also to take into account
the higher Fourier harmonics of the charged fields
(\ref{fihij}), (\ref{XiEtj}) and of the ghosts. After
this, the factor multiplying the derivative term
 in Eq. (\ref{Lkin}) is traded for the expression
 $$1 - \frac {3c_V} {4g} 
\sum_{\vec{n}} \frac 1 {|\vec{A}^{(j)} - 
2\pi \vec{n}/g|^3}\ .$$
The divergent part of this sum gives the renormalization
of the field theory coupling constant with the correct 
coefficient.

\section*{Acknowledgements}
I am deeply indebted to E. Akhmedov, M. Shifman, 
A. Vainshtein, and P. van Baal for illuminating discussions. 
I appreciate warm 
hospitality extended to me at the University of Minnesota, 
where this work was mostly done.

\section*{Appendix A: Calculation of the effective supercharge for
SQED.}
\setcounter{equation}0
\renewcommand{\theequation}{A.\arabic{equation}}

The calculation of the effective supercharges and Hamiltonian for SQED was done in 
Ref.\cite{jaSQED} but was not presented there in enough details, 
so that it was not absolutely trivial even for myself  to reproduce it after 15 years.  We decided to redo the calculations
in the most explicit way. 

To calculate the matrix 
elements in  Eq.(\ref{Qeffpopr}), we need to
know not only the ground state (\ref{vacH0}) of the Hamiltonian $H^{(0)}$, but also the
relevant excited states. The wave functions of the states providing for nonzero
matrix elements $\langle H^{(2)} \rangle_{n0}$ were written in Ref. \cite{jaSQED}:
   \be
\label{excitbos}
 |\vecg{+}\rangle  = \sqrt{2} 
\frac e\pi A [1 - e A(\bar\phi \phi + \bar\chi \chi)]
\nonumber \\
\times \exp\{-eA(\bar\varphi\varphi + \bar\chi\chi) \}
\left[\xi^\alpha \eta_\alpha + \xi^\alpha (\sigma_k)_\alpha^{\ \beta} 
\eta_\beta A_k/A \right]\ , \nonumber \\
 |\vecg{-}\rangle  = \sqrt{2} 
\frac e\pi A^2 (\bar\phi \phi - \bar\chi \chi)
\exp\{-eA(\bar\varphi\varphi + \bar\chi\chi) \}
\left[\xi^\alpha \eta_\alpha + \xi^\alpha (\sigma_k)_\alpha^{\ \beta} 
\eta_\beta A_k/A \right]\ , \nonumber \\
|\vec{l} \rangle  = \sqrt{2} \ .
\frac e\pi A 
\exp\{-eA(\bar\varphi\varphi + \bar\chi\chi) \}
 \xi^\alpha (\sigma_k)_\alpha^{\ \beta} 
\eta_\beta \left( \delta_{lk} - \frac{A_l A_k}{A^2} \right)\ .
  \ee
They are all bosonic and have energy $2eA$. The states $|\vec{l} \rangle $ are not 
all
linearly independent and normalized according to
 \be
\label{normjk}
\langle \vec{l} | \vec{k} \rangle \ =\ \delta_{lk} - \frac {A_l A_k}{A^2} \ .
 \ee 

We can evaluate now the second term in Eq.(\ref{Qeffpopr}). Note first of all that 
the wave function $(e^2/2)(\bar\varphi \varphi - \bar\chi \chi)^2 |0\rangle$ has the 
projection on the vacuum state [irrelevant in the context of Eq.(\ref{Qeffpopr})] 
and also the projection on the state $|\vecg{+}\rangle$,
 \be
\label{plusnol}
\left\langle \vecg{+} \left| \frac {e^2}2 \left(\bar\varphi \varphi - \bar\chi 
\chi\right)^2
\right| \vec{0} \right\rangle \ =\ - \frac 1{2\sqrt{2} A^2}\ ,
 \ee
while all other matrix elements are zero. The result of the action of
the  Laplacian
$\Delta_A$ on the vacuum state is not just a function, but a
differential operator, 
 \be
\label{laplmatel}
\Delta_A |\vec{0}\rangle \ =\ |\vec{0}\rangle \Delta_A + \sqrt{2} \left(|\vecg{+}
\rangle  \frac {A_i}{A^2}
+  |\vec{l}\rangle \frac 1A \right) \frac {\partial}{\partial A_l} + 
|\vecg{+}\rangle \frac {1}{ \sqrt{2}A^2}  \ .
  \ee
Combining Eq.(\ref{plusnol}) and Eq.(\ref{laplmatel}), we arrive at the following
expressions for nondiagonal matrix elements
   \be
\label{matelH2}
\left\langle \vecg{+} \left| H^{(2)}
\right| \vec{0} \right \rangle  &=& - \frac 1{\sqrt{2} A^2} \left(1
+  A_k
 \frac {\partial}{\partial A_k} \right)\ , \nonumber \\
\left\langle \vec{l} \left| H^{(2)}
\right| \vec{0} \right \rangle  &=& - \frac 1{\sqrt{2} A}\frac 
{\partial}{\partial A_l}\ .
   \ee
We need also the matrix elements of the supercharge $Q_\alpha^{(1)}$. They are
 \be
\label{matelQ1}
\left\langle \vec{0} \left| Q^{(1)}_\alpha 
\right| \vecg{+} \right \rangle &=& \frac {iA_k}{2A^2} (\sigma_k)_\alpha^{\ \beta} 
\psi_\beta\ , \nonumber \\
\left\langle \vec{0} \left| Q^{(1)}_\alpha 
\right| \vec{l} \right \rangle &=& \frac {i}{2A} (\sigma_k)_\alpha^{\ \beta} 
\psi_\beta \left( \delta_{lk} - \frac{A_l A_k}{A^2} \right)\ , \nonumber \\
\left\langle \vec{0} \left| Q^{(1)}_\alpha 
\right| \vecg{-} \right \rangle &=& - \frac {i}{2A} \psi_\alpha\ .
 \ee
To evaluate the middle term in Eq.(\ref{Qeffpopr}) we need only two first
matrix elements. The corresponding correction to the effective supercharge is
 \be
 \label{QH2popr}
\delta_{H_2}Q_\alpha^{\rm eff} \ =\ \frac i{4\sqrt{2} e A^3} 
 (\sigma_k)_\alpha^{\ \beta} 
\psi_\beta \left( \frac {\partial}{\partial A_k} + \frac {A_k}{A^2} \right)\ .
  \ee

Let us discuss now the last term in Eq.(\ref{Qeffpopr}). Besides the bosonic
intermediate states (\ref{excitbos}), 
there are also fermionic states providing
for nonzero matrix elements  $\langle H^{(1)} \rangle_{m0}$. 
One of such states has the eigenfunction
  \be
\label{excitferm}
  \phi_\xi^\gamma  = 2 
\frac e\pi A^{3/2} \chi
\exp\{-eA(\bar\varphi\varphi + \bar\chi\chi) \}
\left[\xi^\gamma + \xi^\alpha (\sigma_k)_\alpha^{\ \gamma} 
\ \frac{ A_k}A \right]
 \ee
and there are three others similar. 
The important fact is that all these states have the
same energy $E_{\rm ferm} = 2eA$. And this means that we can do the sum $\sum_m$ in
 Eq.(\ref{Qeffpopr}) without tears and write 
 \be
\label{sumpom}
\sum'_m \langle H^{(1)} \rangle_{nm}
\langle H^{(1)} \rangle_{m0} \frac 1{E_m} \ = \  \frac 1{2eA}
\langle [H^{(1)}]^2 \rangle_{n0} \ .
 \ee
The explicit calculation gives
 \be
\label{matelH12}
 \left\langle \vecg{+} \left| [H^{(1)}]^2
\right| \vec{0} \right \rangle &=&  - \frac {e\sqrt{2}}A\ , \nonumber \\
\left\langle \vecg{-} \left| [H^{(1)}]^2
\right| \vec{0} \right \rangle &=& - \frac {e\sqrt{2}A_i}{A^2} \bar\psi\sigma_i \psi 
\ , \nonumber \\
\left\langle \vec{l} \left| [H^{(1)}]^2
\right| \vec{0} \right \rangle &=&   \frac {ie\sqrt{2} A_k}{A^2} \epsilon_{plk} 
\bar\psi\sigma_p\psi \ .
  \ee
The corresponding correction to $Q_\alpha^{\rm eff}$ is
  \be
 \label{QH1popr}
\delta_{H_1} Q_\alpha^{\rm eff} \ =\ \frac {iA_k}{4\sqrt{2} e A^5}
\left(3 \psi_\alpha \bar\psi \sigma_k \psi -  
 (\sigma_k)_\alpha^{\ \beta} 
\psi_\beta \right) \ .
  \ee
We obtain finally
 \be
\label{Qeffres}
 Q_\alpha^{\rm eff} 
\ =\  \frac 1{\sqrt{2}} (\sigma_k)_\alpha^{\ \beta} 
\psi_\beta \left( 1 - \frac 1{4eA^3} \right) P_k +
 \frac {3iA_k}{4\sqrt{2}e A^5} \psi_\alpha \bar\psi \sigma_k \psi\ .
 \ee
This operator acts on the coefficient $r_0(\vec{A}, \psi_\alpha)$ 
in the expansion (\ref{expan}).
As was explained in Ref.\cite{jaSQED}, the standard normalization 
condition for the 
total wave function bring about the extra factor   
$f^{-2}(\vec{A})$, $f(\vec{A}) = 1 - 1/(4eA^3)$,   
in the normalization condition for $r_0(\vec{A}, \psi_\alpha)$. 
The effective supercharge acting on the 
effective wave 
function with the standard normalization $\Psi^{\rm eff} 
(\vec{A}, \psi_\alpha) = 
f^{-1}(\vec{A})r_0(\vec{A}, \psi_\alpha) $,   is obtained by 
wrapping 
$$  Q_\alpha^{\rm eff}  \longrightarrow  f^{-1}(\vec{A}) 
Q_\alpha^{\rm eff}
f(\vec{A}) $$
and we arrive at the expression 
which coincides, indeed,  with the first line of Eq.(\ref{QHN2}).

\section*{Appendix B: Roots and root vectors for $SU(3)$ }
\setcounter{equation}0
\renewcommand{\theequation}{B.\arabic{equation}}
 
The group $SU(3)$ has rank 2, there are two commuting generators 
$\lambda^3/2$ and
$\lambda^8/2$ ($\lambda^a$ are the standard Gell-Mann matrices). 
There are 3 positive roots.
The root vectors are 
 \be
\label{vecSU3} 
 e_1 &=& \frac 12 (\lambda^1 + i\lambda^2)\ ,\ \ \ 
e_2 = \frac 12 (\lambda^6 + i\lambda^7)\ ,\ \ \
e_3 = \frac 12 (\lambda^4 + i\lambda^5)\ , \nonumber \\  
 f_1 &=& \frac 12 (\lambda^1 - i\lambda^2)\ ,\ \ \ 
f_2 = \frac 12 (\lambda^6 - i\lambda^7)\ ,\ \ \
f_3 = \frac 12 (\lambda^4 - i\lambda^5)\ .
 \ee
The relevant  root forms are 
  \be
\label{formSU3}
\alpha_1(A^s_k) = A^3_k\ , \ \ \ \  \alpha_2(A^s_k) = \frac{-A^3_k + \sqrt{3} A^8_k}2\ , \ \ \ \  
\alpha_3(A^s_k) = \frac{A^3_k + \sqrt{3} A^8_k}2 
   \ee
and similarly for $\alpha_j(\lambda^s_\alpha)$. One can observe that
$\alpha_3 = \alpha_1 + \alpha_2$. The  matrix of scalar
products $c_{jj'}$ is
 \be
\label{Cartan}
c_{jj'} \ =\ \frac 12 \left( 
\begin{array}{ccc} 2 & -1 &  1 \\
- 1 & 2 & 1 \\ 
 1 & 1 & 2 \end{array} \right)\ .
 \ee 
It has an extra factor $1/2$ compared to a more usual definition 
\cite{Slansky}. With Eqs. (\ref{formSU3}) and (\ref{Cartan}) in hand, one can
explicitly check that the relations (\ref{sumroot}), (\ref{sumcal}) hold.


\begin{thebibliography}{40}

\bibitem{Witten}
E. Witten,
Nucl.\ Phys.\  {\bf B202} (1982)  253.

\bibitem{Trieste} A.V. Smilga, in: {\it Proc. Int. Workshop on 
Supermembranes
and Physics in 2+1 dimensions} (Trieste, July, 1989), eds. 
M.J. Duff, C.N. Pope,
E. Sezgin (Worlds Scientific, Singapore, 1987)

\bibitem{deWitt}  B. de Witt, M. L\"uscher, and H. Nicolai, Nucl. Phys. 
{\bf B320} (1989) 135.

\bibitem{Kac2} V.G. Kac and A.V. Smilga, Nucl. Phys. {\bf B571}[PM] 
(2000) 515.

\bibitem{jaSQED} A.V. Smilga, Nucl. Phys. {\bf B291} (1987) 241.

\bibitem{Ivanov} E.A. Ivanov and A.V. Smilga, Phys. Lett. 
{\bf B257} (1991) 79.

\bibitem{vanBaal} P. van Baal, hep-th/0112072.

\bibitem{Lusch} M. L\"uscher, Nucl. Phys. {\bf B219} (1983) 233.

\bibitem{Akhm} E.T. Akhmedov and A.V. Smilga, in preparation.

\bibitem{Zhelob} D.P. \^Zelobenko, {\it Compact Lie groups and their 
representations}, American Mathematical Society, Providence, 1973.

\bibitem{Kac1} E. Witten, J. High Energy Phys. {\bf 9802} (1998) 006;
A. Keurentjes, A. Rosly, and A.V. Smilga, Phys. Rev. {\bf 58} (1998)
081701; V.G. Kac and A.V. Smilga, hep-th/9902029, published in
{\it The Many Faces of the Superworld} (World Scientific, 2000), ed.
M.A. Shifman; A. Keurentjes, J. High Energy Phys. {\bf 9905} (1999) 001, 014.

\bibitem{Slansky} R. Slansky,   Phys. Rept. {\bf 91} (1981) 1.




\end{thebibliography}
\end{document}